\documentclass[letterpaper,aps,preprintnumbers,longbibliography,longbibliography]{revtex4-1}
\usepackage[latin9]{inputenc}
\usepackage{amsmath}
\usepackage{amssymb}
\usepackage{graphicx}
\usepackage{esint}
\usepackage[scr=boondoxo, scrscaled=1.05]{mathalfa}
\allowdisplaybreaks

\makeatletter

\pdfpageheight\paperheight
\pdfpagewidth\paperwidth

\pdfoutput=1

\usepackage{epsfig}
\usepackage{graphics}

\usepackage{mciteplus}
\mciteErrorOnUnknownfalse

\makeatother

\begin{document}

\preprint{FERMILAB-PUB-20-100-A}

\title{Cosmological Constant in Coherent Quantum Gravity}

\author{Craig  Hogan}
\affiliation{University of Chicago and Fermilab}

\begin{abstract}
It is argued that quantum states of geometry, like those of particles, should be coherent on light cones of any size.   An exact classical solution, the gravitational shock wave of a relativistic point particle, is used to estimate  gravitational drag   from coherent energy flows, and  the expected gravitational effect  of  virtual transverse  vacuum energy fluctuations on  surfaces of  causal diamonds.
It is proposed that the appropriately spacetime-averaged gravitational effect of the  Standard Model  vacuum state leads to  the observed small  nonzero value of the cosmological constant,  dominated by gravitational drag of virtual gluonic strings at the strong interaction scale. \end{abstract}
\maketitle

According to the Standard Model of particles and fields, the vacuum is not empty, but is full of fluctuating quantum fields.   A   na\"ive   estimate suggests that  their gravity is repulsive, and  ought to accelerate the cosmic expansion  like a cosmological constant $\Lambda = 8\pi G \rho_{vac} $, where the   density  $\rho_{vac}\sim  p_{max}^4/c\hbar^3$  depends on the maximum virtual momentum  $p_{max}$  of the fluctuating modes.  A cutoff at the Planck scale of quantum gravity, $p_{max}\sim  m_Pc\equiv \sqrt{\hbar c^3/G}$,  leads to a value $\Lambda \sim c^5/\hbar G\equiv t_P^{-2}$,  about 122 orders of magnitude  larger than the observed value.  Clearly, this simple calculation is missing something important about the gravitational effect of virtual quanta\cite{Weinberg:1988cp}.

In this essay, it is suggested that the missing ingredient is the  quantum coherence of geometry,  the nonlocal quantum nature of gravity at separations much larger than the Planck length.    Nobody knows exactly how this works, but it is widely suspected that significant physical effects of quantum gravity are not confined to the Planck  scale\cite{CohenKaplanNelson1999,Ryu:2006bv,Hooft2018,Carney_2019,Giddings:2019vvj,Verlinde:2019ade}, and may even produce measurable macroscopic fluctuations of causal structure \cite{holoshear,Verlinde:2019xfb,PhysRevD.99.063531,Hogan:2019rsn,Hagimoto_2020}.
We sketch here a calculation that shows how geometrical quantum coherence could  profoundly change the spacetime-averaged gravitational response to quantum fluctuations of the standard quantum vacuum, and  explain a cosmological constant with the observed value.

To start with, recall  that a quantum particle  exists nonlocally, that is, without a definite relationship to space and time.  A particle localized  in space is in a state with delocalized momenta in all directions, which means that its location at another time is indeterminate.  A particle traveling in a definite direction exists with no definite location in the normal plane.
Since gravity and space-time geometry depend on mass-energy in an indeterminate superposition of locations,  the quantum nonlocality, indeterminacy and coherence of matter states must somehow extend to  quantum  states of gravity and space-time geometry on all scales\cite{10.2307/3301037,RevModPhys.33.63}.

 Physical consequences of geometrical quantum coherence follow from correspondence with classical gravity.  Consider a pair of photons emitted in opposite directions by annihilation of a positronium atom. Their directions are exactly opposite to each other  in the frame of the emitting particle, but the axis is indeterminate: 
the  wave function is  a thin shell, a superposition of all directions.  The direction of the photons' axis can be  indeterminate  for an indefinite length of time, extending everywhere on the light cone from the emission event.
The causal structure   remains in a superposition  until the photon state ``collapses'' onto a particular axis.
A consistent, coherent quantum geometry  has to match the active gravity of the photonic energy, wherever it ends up.\footnote{This hypothesis differs physically from the quantized linear field theory of gravity,  where quantum coherence is attached to the amplitude of a plane wave mode whose space-time configuration is defined by an unquantized background geometry. Modes that localize a point particle in the plane normal to its  axis of propagation are not independent in coherent quantum geometry, since they are entangled with  spherical  surfaces of causal structure in all directions around an event. As shown here, this radically changes the  space-time-averaged gravitational effect of vacuum fluctuations in energy-momentum.}

Gravity can be consistent with the  outcome of any particle experiment   if the light cone is itself a coherent quantum object, like the wave function of the  photons. In that case,  a  measurement  nonlocally affects  the causal  configuration of an extended region of  space and time; the particles  entangle with the space-time, in much the same way that Schr\"odinger's decayed atom  entangled with his macroscopic, poisoned cat.
Quantum space-time is then fundamentally discontinuous, nonlocal and discrete.  
Classical space and time  emerge as a statistical average of many elements,  similar to the way a  quasi-continuous fluid  emerges from a  discrete molecular system\cite{Jacobson1995,Jacobson:2015hqa,Padmanabhan:2013nxa}.

\begin{figure}[t]
\begin{centering}
\includegraphics[height=3in]{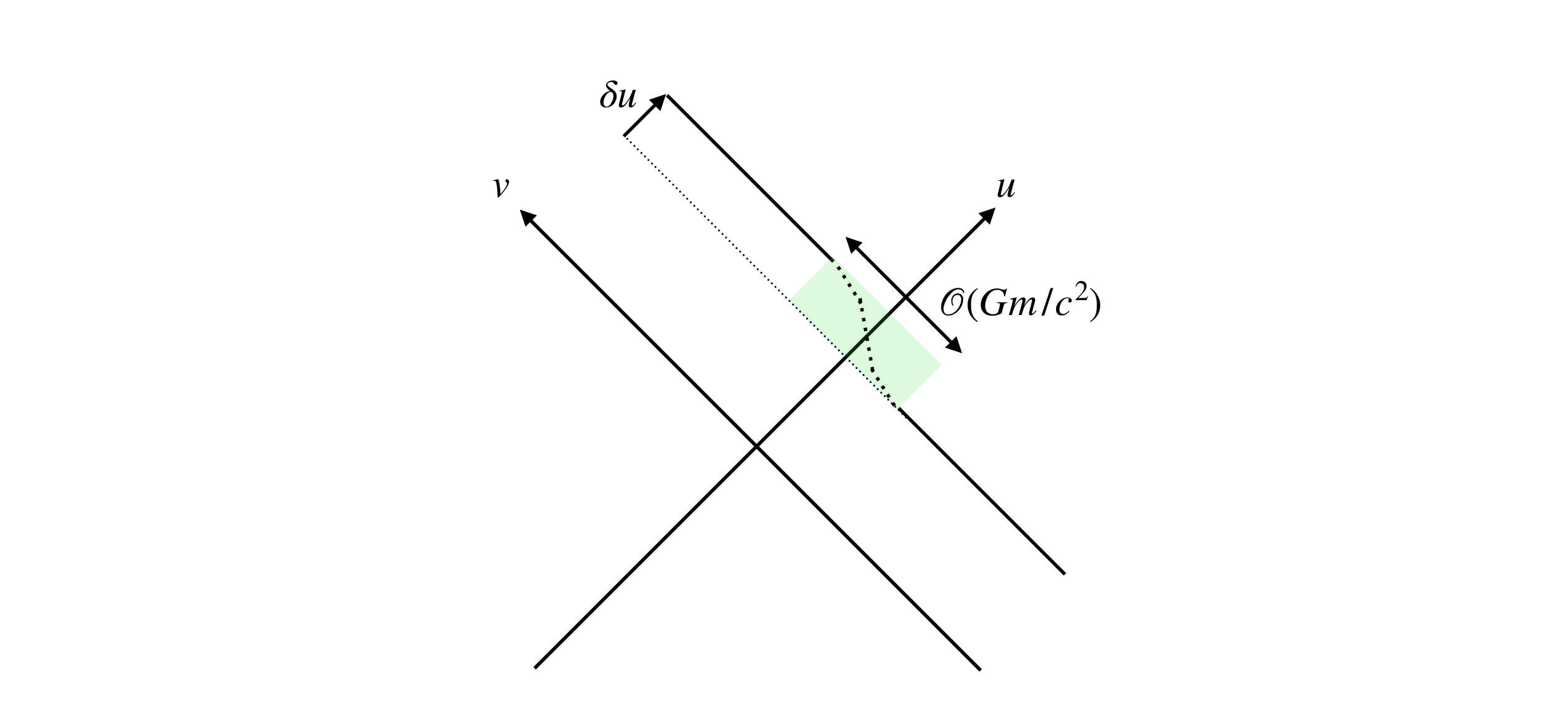}
\par\end{centering}
\protect\caption{ The classical gravitational  dislocation of causal structure  from a  very fast-moving point mass $m$ \cite{Aichelburg1971,DRAY1985173,Hooft:2016itl}. 
In  null coordinates $u,v$ on the light cone of the particle moving in the $u$ direction with momentum $p>>mc$, 
the gravitational drag of the particle  (Eq. \ref{drag})  creates a  displacement $\delta u$ that varies on opposite sides of the trajectory.  The physical displacement is axially symmetric around the particle trajectory and  independent of the impact parameter of the particle, $x_\perp$.   \label{shock}}
\end{figure}

To quantify the physical effect of geometrical coherence, consider first the classical gravitational shock wave  of  a localized relativistic point particle in flat space-time \cite{Aichelburg1971,DRAY1985173,Hooft:2016itl}.
A  particle with momentum  $p$ and impact parameter  $x_\perp$ creates a coherent  displacement   everywhere on a conical light sheet  (Fig. \ref{shock}),
\begin{equation}\label{drag}
\delta u = 4 G p c^{-3}.
\end{equation}
This displacement represents a real, physical distortion of causal structure; the motion of the particle ``drags'' the space-time along with it.
The shock creates an  axially symmetric distortion of the causal structure  in the  directions normal to the particle axis. 

  \begin{figure}[t]
\begin{centering}
\includegraphics[width=\linewidth]{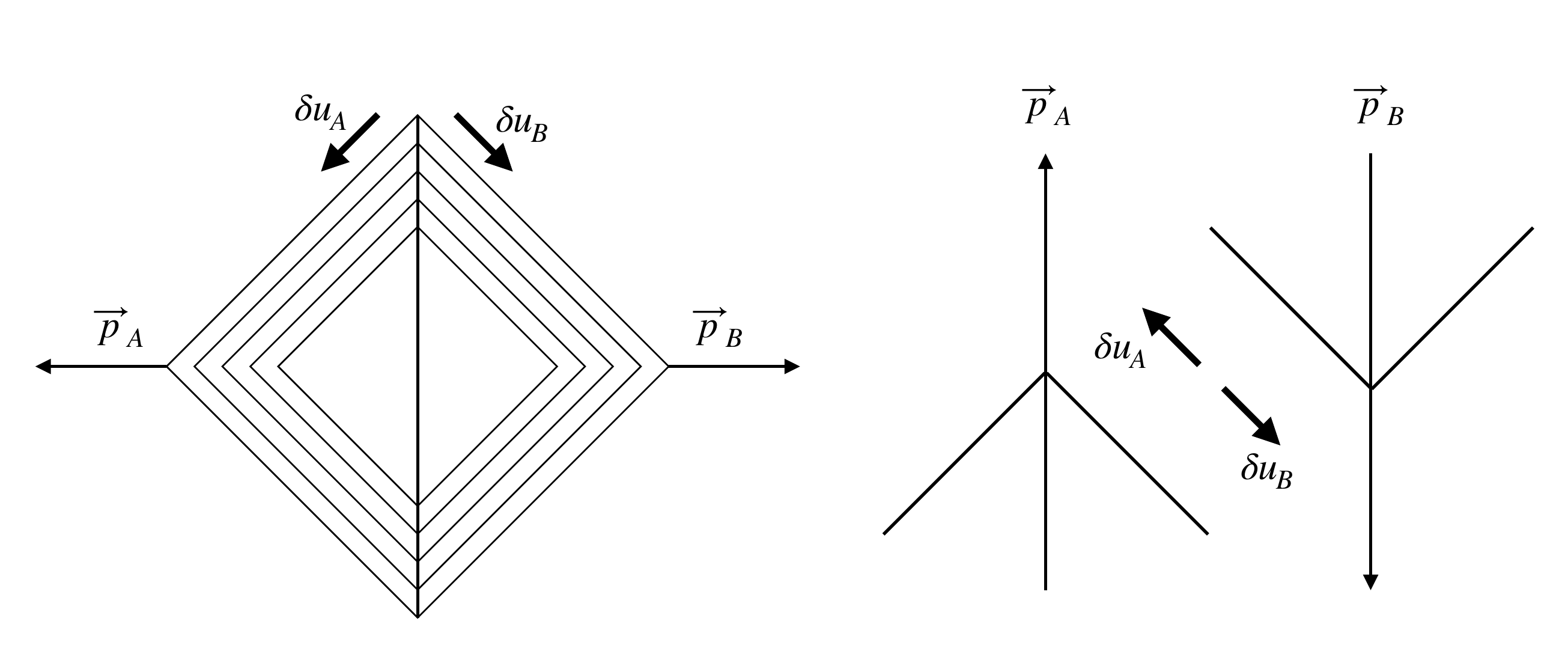}
\par\end{centering}
\protect\caption{ Coherent spatial patterns of conical shock waves from  counterpropagating particles, with  perturbations $\delta u_A, \delta u_B$ due to frame dragging.  At left,  an inwards shock forms  from a pair created at the origin by annihilation, shown at a  series of successive times. The orientation of a  pair of entangled particles can  remain indeterminate to arbitrarily large scales;  causal structure is placed in a superposition on the whole series of  causal diamond surfaces far from the particles,  with a compression in the plane normal to the axis.    At right, two particles travel transversely to an observer in between. In a quantum vacuum,  a nonlocally correlated state  with fluctuating angular momentum can cause tiny rotational fluctuations by dragging the local inertial frame (Eq. \ref{acceleration}), and a net outwards acceleration of geodesics (Eq. \ref{cosmic}).
 \label{shocks}}
\end{figure}

This classical solution allows us to  construct simple examples of  coherent  quantum states of geometry associated with coherent energy flows of particle states. For  a state like our positronium annihilation,  with two counterpropagating particles in a superposition of orientations  (Fig.   \ref{shocks}), the space-time is also in a superposition of  spacelike distortions. The  orientation of the distortion is entangled with the particle axis, and extended over space and time.
 A collapse onto a definite particle trajectory leads to  collapse of an extended  causal diamond, with a dislocation in the normal directions, wherever they end up.
 For two coplanar classical  particle trajectories $A,B$ with $\vec p_A=-\vec p_B$ that do not pass through a point, but are separated by an impact parameter $x_{AB}$ as they pass by, gravitational drag  creates a rotational   impulse,   a  ``twist'' of the local inertial frame.   Setting aside geometrical factors of the order of unity, the  angular velocity of the twist normal to the plane is about
\begin{equation}\label{twist}
 \omega \sim c \delta u_{AB}/x_{AB}^2\sim Gp_{AB}/c^2x_{AB}^2,
\end{equation}
 where  $p_{AB}=|\vec p_{A\perp}-\vec p_{B\perp} |$ denotes the  transverse momentum difference of the particles.
 
We can  use these states  to estimate  the mean gravitational effect of  quantum field vacuum fluctuations on a coherent geometrical state.
The mean rotation $\langle \omega\rangle $ of the inertial frame vanishes,  but there are nonvanishing fluctuations, determined by  the  mean  square transverse  momentum flow on light cones :
\begin{equation}\label{acceleration}
 \langle \omega^2\rangle \sim G^2 c^{-4} \  \langle  p_{AB}^2/x_{AB}^4\rangle_{CD},
\end{equation}
where $\langle  \rangle_{CD}$ denotes an expectation value over  invariant spherical causal diamond surfaces for a world line interval, and $p_{AB}$ denotes the net transverse momentum flow  at opposite points $A$ and $B$ on each surface, with spacelike separation $x_{AB}$. 

The sign of the mean square  $\langle \omega^2\rangle$ is always positive, so the
 fluctuating  impulses  systematically  accumulate outwards displacements of causally-defined surfaces.
Let us suppose that these exotic  fluctuations of the inertial frame  have the same  effect on the radial component of geodesics  in the time-averaged, emergent geometry as classical rotation,  that is to say,  a mean centrifugal  acceleration  $\ddot r$   proportional to  separation $r$:
 \begin{equation}\label{cosmic}
  \langle  \ddot r/r \rangle= \langle \omega^2\rangle.
\end{equation}
The  time-averaged physical effect on freely falling trajectories is then the same  as a cosmological constant $\Lambda= 3\langle \omega^2\rangle $:
the gravity of  fluctuating twists in the  virtual vacuum pulls  the emergent space-time apart.


The gravitational effect from the particle vacuum with this kind of coherent quantum  average over causal diamonds is much less than the na\"ive calculation outlined above based on uncorrelated field modes. The  sum of linear  dislocations from  virtual pairs in all directions averages to zero, so   approximately-pointlike virtual particle states  produce  approximately-vanishing gravitational fluctuations.
However,  the gravitational effect does not exactly vanish. In the Standard Model,  the pointlike-particle approximation breaks down from the strong  self-coupling of gluons\cite{Wilczek1999,Wilczek:2000ih} 
 at the QCD  energy scale $m_{Q}c^2$,  which largely determines the masses of protons, neutrons, and atoms, as well as the range of nuclear interactions.
Coherent gluonic fluctuations at this scale   include  significant spacelike correlations of transverse momentum.

In a coherent quantum geometry, these  momentum fluctuations produce a tiny bit of gravitational repulsion.
The  gravitational effect can be estimated from  Eq. (\ref{acceleration}), with $ p_{AB}/c\sim \hbar/x_{AB}c\sim m_{Q}$:
\begin{equation}\label{estimate}
\langle \omega^2\rangle
\sim  G^2 m_{Q}^6 c^2/\hbar^{4}
  \sim  t_P^{-2} (m_{Q}/m_P)^6,
\end{equation}
where  
$t_P\equiv \sqrt{\hbar G/ c^5}= 0.54 \times 10^{-43}{\rm sec}$ and
$m_P\equiv \sqrt{\hbar c/G}= 1.22\times 10^{19} {\rm GeV}/c^2$.
The measured cosmic acceleration\cite{Tanabashi:2018oca},
\begin{equation}\label{rate}
\sqrt{\langle \omega^2\rangle} = \sqrt{\Lambda/3} \approx 1.0 \times 10^{-61} t_P^{-1},
\end{equation}
would be produced by   vacuum fluctuations  with  mass  on the order of
\begin{equation}
m_{Q}\sim ( \sqrt{\langle \omega^2\rangle} t_P)^{1/3} m_P \sim 0.5\times 10^{-20} m_P \sim 60\ {\rm MeV}/c^2.
\end{equation}
This rough  estimate  is remarkably close to  the actual scale of strong interactions, as  measured  by the masses of their lowest-energy real particle states, the  pions\cite{Tanabashi:2018oca}:
\begin{equation}
m_{\pi^0}=  135\  {\rm MeV}/c^2.
\end{equation}


Like pions and other mesons,  vacuum fluctuations  on the nuclear scale can be visualized not as virtual pointlike particles, but  as  spatially extended strings  of gluon condensate with quarks at the ends.  The bulk of the stress-energy is in the gluonic string, whose stress-energy is dominated by  negative pressure or tension.  The gravity of a string, unlike that of a point particle, is repulsive,  so  stringlike virtual-gluon states naturally lead  to a small cosmic acceleration, with a value similar to that observed.\footnote{Similarly, in a classical thermodynamic view,  the  adiabatic  ``work''  done on virtual gluonic strings slightly stretched by gravitational repulsion  is the source of the new vacuum ``energy'' in the expanded space. This microscopic view of the cosmological constant has a complementary holographic interpretation in terms of geometrical information: the entanglement entropy of the horizon matches the independent field degrees of freedom below the QCD scale\cite{Hogan:2015b}.  It  differs physically from  other recent microscopic models\cite{Perez:2017krv,Perez:2018wlo} seeking to connect the cosmological constant with standard model vacuum states,  based on Planck scale discreteness and electroweak scale  torsion.}

This scenario  roughly accounts  for the well-known puzzling  coincidence\cite{Z68,Hogan:1999wh,Hogan:2015b} of  cosmic acceleration  (Eq. \ref{rate}) with the rate of stellar evolution, which  depend on the same power of the  small number $m_{Q}/m_P$.  The origin of this small dimensionless number (and the relative weakness of gravity) is well understood in the Standard Model,  from the logarithmic running of the strong coupling constant with energy scale\cite{Wilczek1999}.

The departure of the cosmological constant from zero in this picture is  a low-energy, infrared phenomenon. Its physical value depends  only on Standard Model fields, and a standard gravitational response to an appropriate  average over their quantum stress-energy fluctuations.  The expected  gravitational effect of  vacuum fluctuations (Eq. \ref{acceleration}) depends on nonlocal  spacelike correlations of  QCD fields, which can in principle be calculated on a lattice, as nucleon masses are\cite{Wilczek:2000ih,Tanabashi:2018oca}.  It does not depend on  detailed properties of Planck-scale physics, except for simple symmetries of  coherent geometrical states on scales $>>t_P$.

  \begin{acknowledgments}
This work was supported by the Department
of Energy at Fermilab under Contract No. DE-AC02-07CH11359.  
\end{acknowledgments}

\break
\bibliography{CCbib}

\end{document}